PAPER TYPE (ARTICLE)

# Reverse Energy Flows in Two-Dimensional Photonic Crystals and Analogies with Vortex Formation and Analogous Flows in Hydrodynamics

**Andrey Pryamikov***

Prokhorov General Physics Institute of the Russian Academy of Sciences, Moscow, 119991, Russia
*Corresponding Author: Andrey Pryamikov. Email: pryamikov@mail.ru

**ABSTRACT:** This paper examines the connection between photonic band-gap formation in two types of two-dimensional photonic crystals and the emergence of reverse electromagnetic energy flows generated by linearly polarized plane waves incident on a photonic-crystal slab. We show that these reverse energy flows, observed in both transmitted and reflected fields, originate from vortex structures in the Poynting vector. The resulting energy-flow patterns exhibit striking analogies to vortex formation in fluid motion past obstacles. The geometry and dynamics of the Poynting-vector vortices determine whether the incident electromagnetic energy is impeded, leading to the formation of photonic band gaps, or instead guided through the structure, enabling transmission.

---

## 1 Introduction

Electromagnetic energy flows is a fundamental dynamical characteristic of electromagnetic radiation and it is quantified by the Poynting vector. Electromagnetic energy backflow is a phenomenon that occurs when the direction of the Poynting vestor is opposite to that of the wave field propagation. Situations where the Poynting and wave vectors have opposite directions are well known in the case of metamaterials. While it seems obvious that the energy flow of a light beam in free space has the same direction as the beam propagation direction then the occurrence of a reverse energy flows or energy backflows when the Poynting vector has negative values within some regions, may seem counterintuitive. In addition, it is known that in classical fluid dynamics, the flow of fluid is characterized by velocity vectors that describe the direction and magnitude of fluid motion. When the flow encounters obstacles or particular boundary conditions, complex structures such as vortices—regions of swirling flow—can develop. These vortices can cause reverse flow regions where the fluid moves contrary to the main flow direction, significantly influencing transport phenomena and energy distribution. Similarly, in the context of electromagnetic wave propagation within photonic crystals, the Poynting vector represents the directional energy flux density of the electromagnetic field. The formation of Poynting vector vortices corresponds to localized regions where the energy flow circulates, creating reverse energy flow analogous to fluid flow reversal in vortices. This behavior impacts the overall transmission and reflection of waves, just as vortices in fluids affect flow resistance and transport.

In electrodynamics, the conditions for a negative Poynting vector to arise and the energy backflow effect to emerge require complicated nonparaxial electromagnetic fields. Thus, the effect associated with the formation of reverse energy flows in photonics can be achieved for vector Bessel beams, tightly-focused optical beams in the focal region and in other similar cases [1 - 6]. It used to be believed that the main applications of energy backflows in photonics are related to microparticle manipulation like 'tractor beams', etc. [7]. However, in a paper from the year 2000 it was demonstrated that reverse flows of electromagnetic field energy could also arise in photonic crystal waveguides [8]. The waveguide was formed by creating a defect in a two-dimensional photonic crystal with circular dielectric rods arranged on a triangular lattice in a background of air. It was shown that, unlike a planar waveguide, in which the Poynting vector of the core mode moves in a straight line, in a photonic crystal waveguide, there is a twisting in the defect mode energy flows near the boundaries of the transmission bands. The resulting reverse energy flows can practically stop the movement of the energy of the defect mode along the waveguide axis [8]. A similar phenomenon occurs in hollow - core fibers with a curved core-cladding boundary. In this case, there is no translational symmetry in the distribution of air holes or dielectric rods as in a 2D photonic crystal, but nevertheless, reverse energy flows of the air - core modes are formed in the walls of the fiber cladding boundary [9]. In addition, there are quite a few works devoted to analogies arising between electromagnetism and hydrodynamics. In particular, the hydrodynamic approach to the description of optical vortices was covered in [10 -12], and general problems arising when drawing analogies between electromagnetism and hydrodynamics were discussed in [13 - 17].

In this paper we consider the issue of the relationship between the effective reflection of plane waves from 2D photonic crystal slabs and the formation of reverse flows of radiation energy arising in this case. It was demonstrated that when reflecting radiation in the photonics band gap mode, the structure of the reverse energy flows of the electromagnetic field leads to a significant decrease in the total energy flow going through the photonic crystal. Similar phenomena can occur in hydrodynamics, for example in the well-known Tesla valve invented in 1920 by Nikola Tesla (Fig. 1), when asymmetric forward and reverse fluid flows meet in the same channel. Valves are structures with a higher-pressure drop when fluid flows in the reverse rather than forward direction. Tesla valve is a series of interconnected, asymmetric, tear-shaped loops (islands and bends) that allow it to move fluid in a single direction without any moving parts. This device generates a substantial resistance difference between two flow directions. When fluid flows in the desirable direction, it moves smoothly through the channels with minimal resistance. However, for the flow in the reverse direction, the geometry of the valve creates vortices disrupting the flow and increasing the resistance to the backward leakage. Analogies with the Tesla valve are used in various fields of science [18].

Thus, conducting research into the formation of photonic band gaps and the associated reverse flows of electromagnetic energy in photonic crystals helps to draw intriguing conceptual parallels with classical fluid dynamics, particularly in the study of flow patterns, vortices, and energy transport.

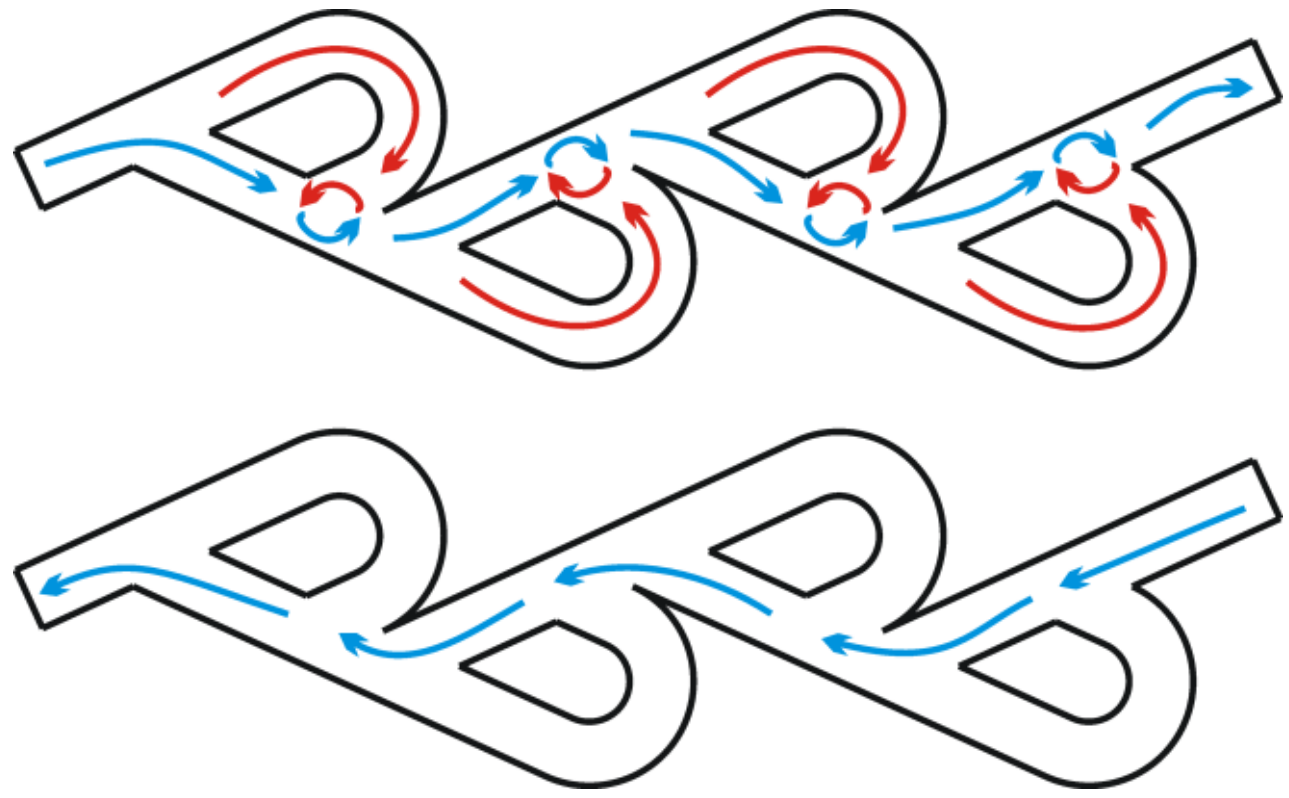

Fig. 1. Schematic representation of the operating principle of the Tesla valve operating principle.

## 2 Reverse energy flows in 2D photonic crystal with air holes

Let us consider the incidence of TE and TM polarized plane waves on a finite two-dimensional photonic crystal with parameters described in the classical work [19]. The square lattice photonic crystal consists of air holes with an air-filling fraction of $f = \pi r^2 / a^2 = 65\%$, where $a$ is the length of the unit cell and $r$ is the radius of the air hole of the photonic crystal. The refractive index of the surrounding material is 4. To calculate the parameters of photonics crystal required to obtain photonic band gaps the normalized frequency values of $\omega a / 2\pi c$ from [19] will be used for both types of polarization, where $\omega = 2\pi c / \lambda$ and $\lambda$ is the operating wavelength of 1.55 μm. The photonic crystal slab consists of ten layers of air holes (Fig. 2) (the *x*-axis is horizontal). TE or TM polarized plane wave is incident on the photonic crystal from the left. The calculations of fields and energy flows were performed in Comsol Multiphysics and scattering boundary conditions were set on the left and right boundaries of the computational domain, which are used in this package when calculating the optical properties of photonic crystals. These boundary conditions are introduced when it is necessary to truncate the computational domain without introducing reflections that could interfere with the results. Scattering boundary conditions simulate an open boundary for wave propagation, making a boundary transparent to outgoing scattered waves, especially at normal incidence. Boundary conditions of the perfect magnetic conductor (TE polarized plane wave) and perfect electric conductor (TM polarized plane wave) type were imposed on the upper and lower boundaries of the computational domain. To obtain good accuracy in calculations, the size of the mesh element was taken to be approximately 20 times smaller than the wavelength used.

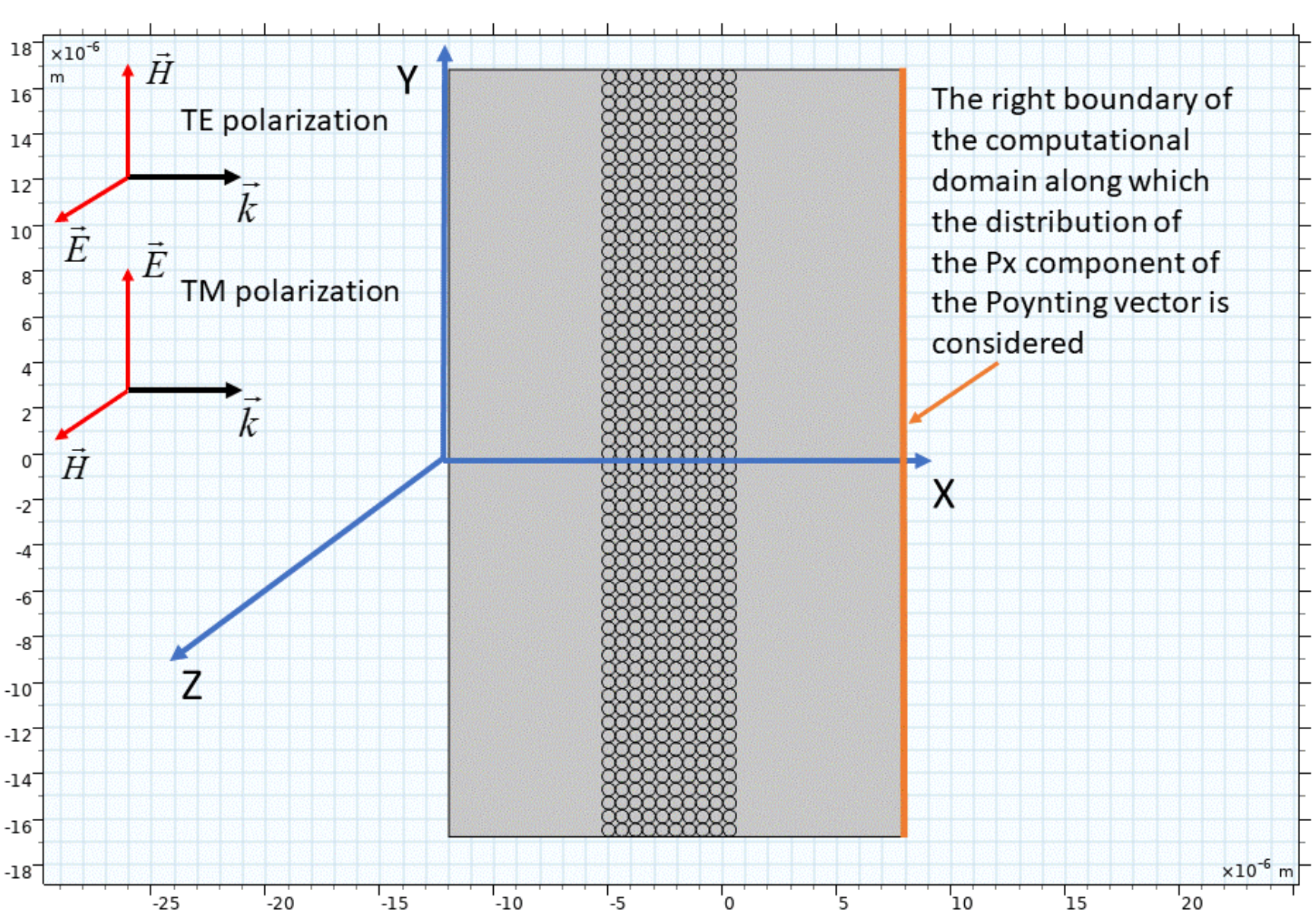

**Figure 2:** A two-dimensional photonic crystal of finite thickness on which TE and TM polarized plane waves fall from the left.

According to the results in [19], the normalized frequency values of 0.38 for the TM polarized incident plane wave (the electric field is directed along the vertical axis Y (Fig. 2)) and for the TE polarized wave (the electric field is directed along Z-axis (Fig. 2)) lie in the photonic band gap region (Fig. 3). The normalized frequency value of 0.5 is outside the photonic band gap for both polarization of incident plane waves, as it can be seen from Fig. 3.

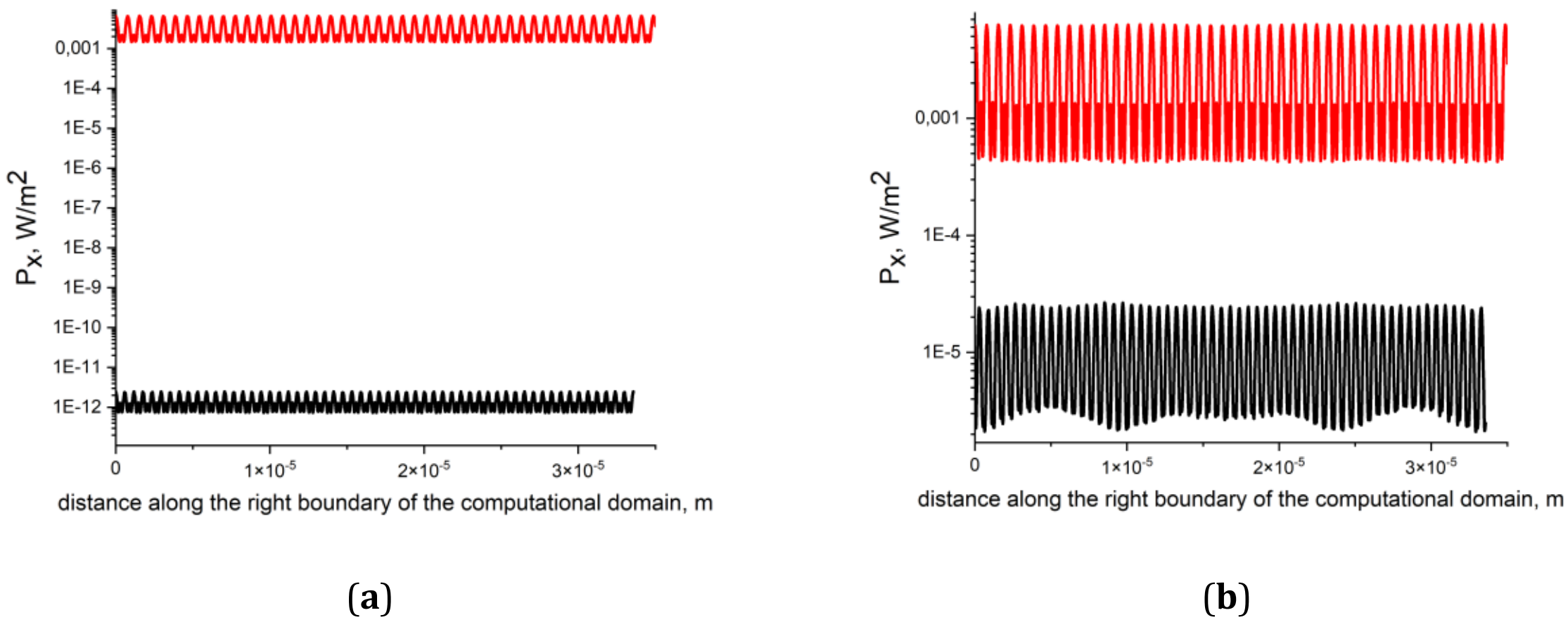

**Figure 3:** (**a**) Distribution of the $P_x$ component of the Poynting vector along the right boundary of the computational domain (Fig. 2) for TM incident polarized plane wave, and a normalized frequency value of 0.5 (red) and 0.38 (black); (**b**) the same distributions for TE polarized incident plane wave, and a normalized frequency value of 0.5 (red) and 0.38 (black).

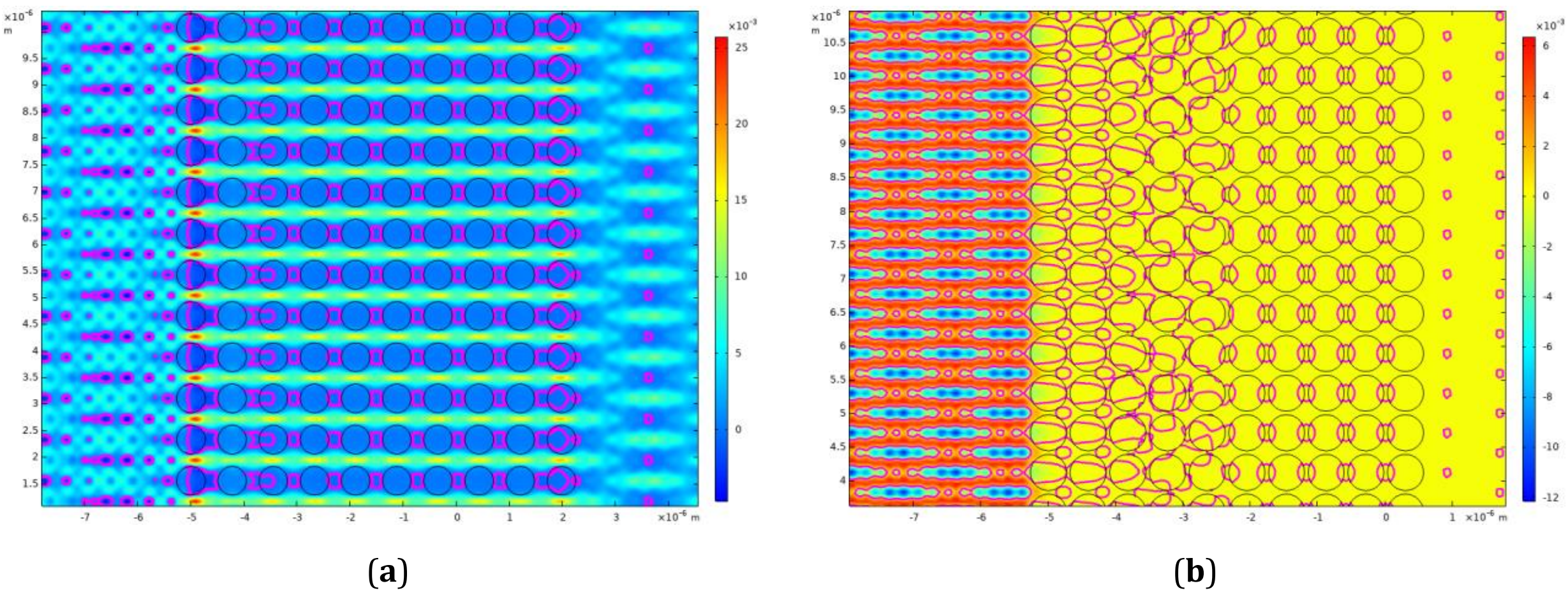

**Figure 4:** (**a**) Distribution of the $P_x$ component in a limited region of a photonic crystal for TM polarized incident plane wave and a normalized frequency value of 0.5 (Fig.3a); (**b**) Distribution of the $P_x$ component in a limited region of a photonic crystal for TM polarized incident plane wave and a normalized frequency value of 0.38 (Fig.3a). The magenta curves are described by the equation $P_x(x, y) = 0$. A plane TM wave is incident on a photonic crystal slab from the left (Fig. 2).

Let us consider the process of formation of photonic band gaps via the energy flows behavior of the plane wave incident on a 2D photonic crystal slab. It is obvious that to characterize the energy flows both inside and outside photonic band gaps, it is necessary to consider the the Poynting vector $P_x$ component (Fig. 2). Fig. 4 and 5 shows the distributions of the $P_x$ component in a limited region of the photonic crystal slab for TM and TE polarized incident waves. As shown in Fig. 2, they fall on the photonic crystal slab from the left. The magenta curves in Fig. 4 and 5 correspond to the equation $P_x(x, y) = 0$. All values of the Poynting vector in all figures in this paper are normalized to the value of the integral of its $P_x$ component along the right boundary of the computational domain (Fig. 2) divided by the length of this boundary for a plane wave propagating in free space .

As it can be seen from Fig. 4 and 5, the interference of the incident and reflected waves leads to the formation of curves $P_x(x, y) = 0$, which have a closed shape and limit the areas inside and outside the photonic crystal where the sign of the $P_x(x, y)$ component is negative. Thus, when reflecting the radiation of an incident plane wave from a photonic crystal, reverse energy flows always arise, moving in the opposite direction to the incident wave. Reverse energy flows exist both inside and outside the photonic crystal slab. Outside the total photonic band gaps, the reverse energy flows occupy small areas of the photonic crystal cross-section between its air holes (Figs. 4a and 5a) and the radiation effectively propagates in the positive direction of the *x*-axis between the rows of air holes (Fig. 4a and 5a). In the case of photonic band gaps, the reverse energy flows occupy significant areas inside the photonic crystal slabs and, most importantly, outside them (Figs. 4b and 5b) thus effectively blocking energy flows in the positive *x*-axis direction (Fig. 2). In this case, the structure of the distribution of reverse energy flows outside the photonic crystal slab for TE and TM polarized incident plane waves is qualitatively similar (Fig. 4b and 5b). This means that the proportion of radiation energy moving in reverse flows is significantly higher in the spectral regions related to photonic band gaps.

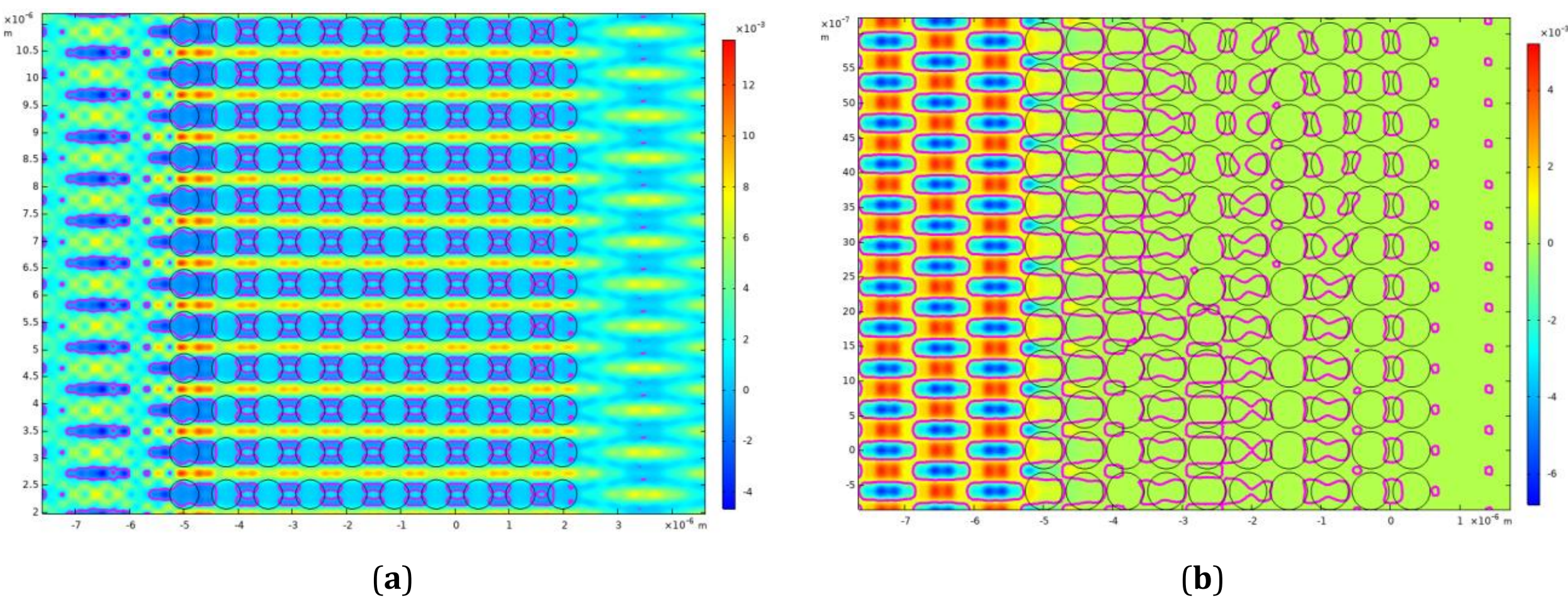

(**a**) (**b**)

**Figure 5:** (**a**) Distribution of the $P_x$ component in a limited region of a photonic crystal for TE polarized incident plane wave and a normalized frequency value of 0.5 (Fig.3b); (**b**) Distribution of the $P_x$ component in a limited region of a photonic crystal for TE polarized incident plane wave and a normalized frequency value of 0.38 (Fig.3b). The magenta curves are described by the equation $P_x(x, y) = 0$. A plane TE wave is incident on a photonic crystal slab from the left (Fig. 2).

As in the case of hollow - core fibers with negative curvature of the core-cladding boundary [9], reverse flows of radiation energy must arise due to the presence of vortex motions of the Poynting vector. Since the Poynting vector is a vector, when all its components are simultaneously zero at a given point in space, the direction of the Poynting vector becomes undefined and a singularity arises. So a Poynting vector vortex may form around this singularity. In our case, the Poynting vector has one component $P_x$ for both the TM polarized incident wave ( $P_x = 1/2\left(E_y H_z^*\right)$ ) and the TE polarized incident plane wave ( $P_x = 1/2\left(-E_z H_y^*\right)$). In case of the complex interference of incident and reflected waves from a photonic crystal slab, the phase fronts of the electric and magnetic fields are distorted, leading to the $P_y$ component of the Poynting vector to arise. The intersection of the $P_x$ and $P_y$ components at any point causes a vortex to form which, in turn, generates a reverse energy flow both along the x-axis and along the y-axis. The vortices centers are the points with coordinates $\left(x_0; y_0\right)$ at which the following condition is satisfied $P_x\left(x_0, y_0\right) = P_y\left(x_0, y_0\right) = 0$. Let us consider, for example, the distribution of curves $P_x(x, y) = 0$ and $P_y(x, y) = 0$ for the case of a TM polarized incident plane wave (Fig. 4). The calculation results are shown in Fig. 6, where the magenta color indicates the curves $P_x(x, y) = 0$ and the black color indicates the curves $P_y(x, y) = 0$.

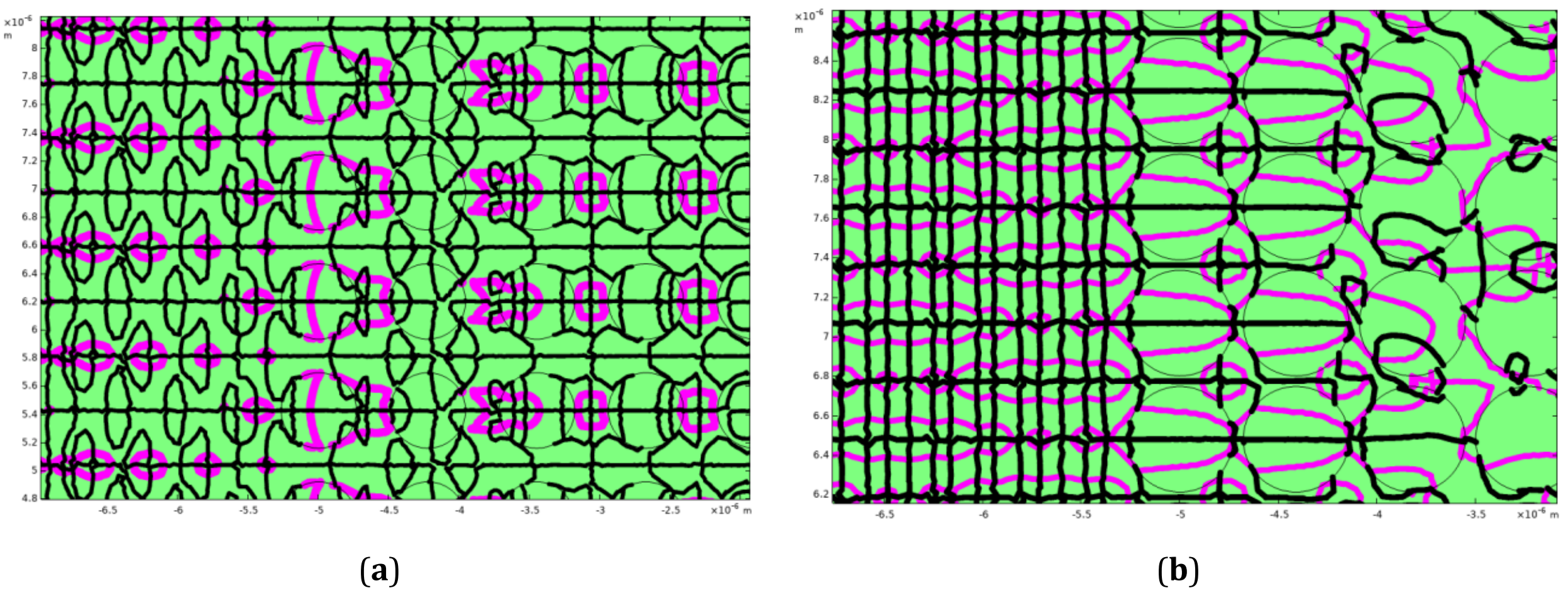

(**a**) (**b**)

**Figure 6:** (**a**) Curves $P_x(x, y) = 0$ (magenta) and $P_y(x, y) = 0$ (black) for the case shown in Fig. 4a. The intersection points of these curves are the centers of the Poynting vector vortices; (**b**) the same curves for the case shown in Fig. 4b.

It is clear from Fig. 6 that it is at the intersection points of these curves that the reverse flows of radiation energy are formed (Fig. 4). Thus, it is the singularities and the corresponding vortices of the Poynting vector of the transmitted and reflected waves that are the cause of the formation of reverse energy flows in the considered photonic crystal slabs with air holes. As it can be seen from Fig. 6, the arrangement of vortices in the cross section of the photonic crystal is directly related to the modes of weak or strong radiation transmission through the photonic crystal slab.

As was indicated in the Introduction, the formation of Poynting vector vortices in 2D photonic crystals may be analogous to vortices formed in a moving fluid flow. The photonic band gap acts as a kind of "barrier" or "forbidden zone" for certain frequencies, analogous to a fluid dynamic system where flow is blocked or redirected by obstacles or changes in boundary conditions. Energy can be carried in fluids by pressure variations or swirling motion and vortices can create areas of low pressure at their cores and higher pressure around them, which disrupts the smooth, forward movement of the liquid and slows down its flow. Within the band gap, the emergence of Poynting vector vortices leads to the blocking or localization of electromagnetic energy flow, much like how fluid vortices can inhibit or redirect fluid motion. Let us consider the distribution of the Poynting vector streamlines (Fig. 7) for the case of photonic crystal slabs shown in Fig. 4.

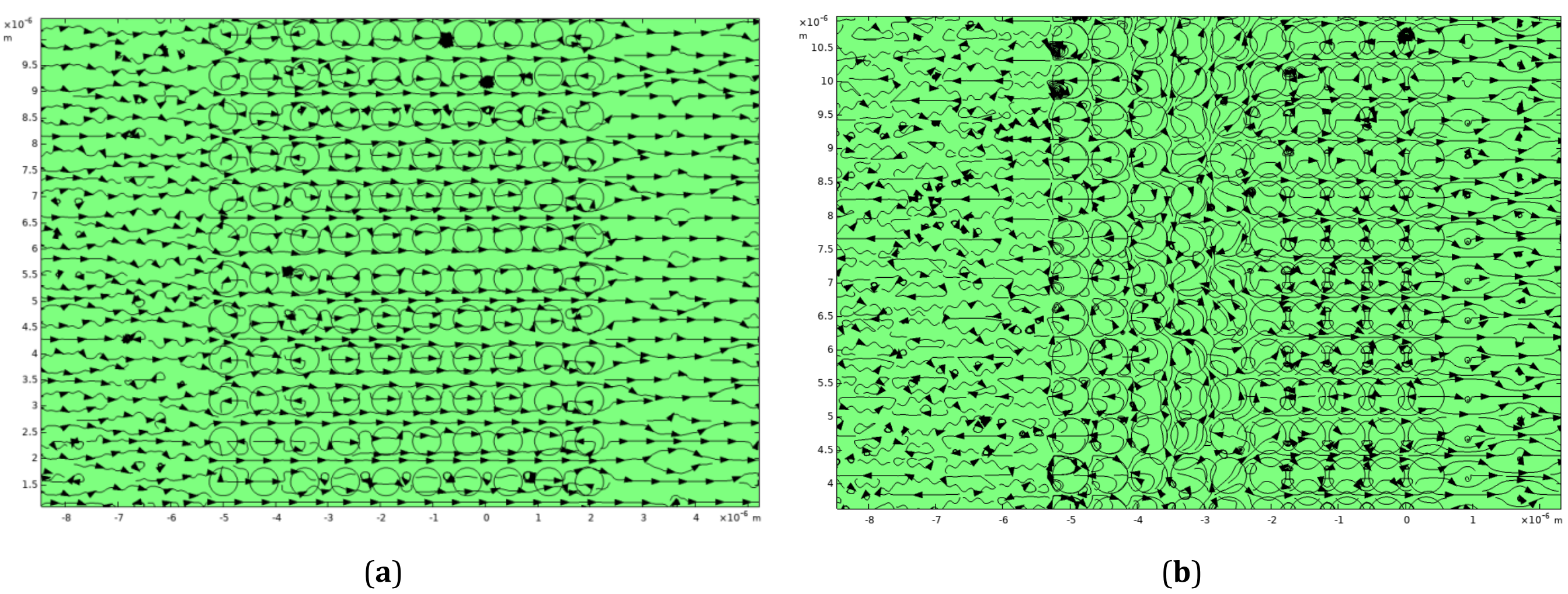

(**a**) (**b**)

**Figure 7:** (**a**) Streamlines of $\vec{P}(x, y) = \{P_x; P_y\}$ for TM polarized incident wave (Fig.4a); (**b**) Streamlines of $\vec{P}(x, y)$ for TM polarized incident wave (Fig.4b).

In the case of strong radiation transmission of the incident TM polarized plane wave, the streamlines of the Poynting vector (Fig. 4a), although deviating from the rectilinear path in the photonic crystal slab, do not form vortices and reverse energy flows in the regions between the air holes (Fig. 7a). Therefore, the energy of the electromagnetic wave is transferred freely through these channels inside the photonic crystal slab. While in the photonic band gap (Fig. 4b) the streamlines of the Poynting vector twist, forming reverse energy flows in the regions between the air holes of the photonic crystal, thereby reducing the radiation transmission of the incident wave through the photonic crystal slab (Fig. 7b). Thus, this behavior is analogous to that of fluid dynamic systems in which fluid vortices can inhibit or redirect fluid motion. In our case, the Poynting vector vortices formed between the rows of air holes and outside the photonic crystal slab perform the same function for the energy flow of the incident plane wave.

To make the analogy between fluid motion in hydrodynamics and energy flow in 2D photonic crystals clearer, let us consider a quantity called vorticity. In a steady plane motion of a fluid, the velocity of a particle $\vec{u}$ is a function of two coordinates $\vec{u}(x, y)$. Based on the velocity field, it is possible to establish such an important value as vorticity, which is a fundamental hydrodynamic vector field and is defined as $\vec{\omega} = \vec{\nabla} \times \vec{u}(x, y)$. As is well known, the elementary movement of a liquid or gas particle generally consists of three components: translational motion, rotation, and deformation of the particle. Rotational motion is called vortex motion, and therefore vorticity is the local angular rate of rotation (spin) of fluid particles. Thus, the condition for the existence of vortices is $\vec{\omega} \neq 0$. The occurrence of vortices must be caused by special reasons, for example, the viscosity of the fluid. In hydrodynamics, the transport of vorticity is governed by the vorticity equation, which is obtained from the Navier-Stokes equation [20]:

$$\frac{\partial \vec{\omega}}{\partial t} + \left(\vec{u}\vec{\nabla}\right)\vec{\omega} = \left(\vec{\omega}\vec{\nabla}\right)\vec{u} + \nu\vec{\nabla}^2\vec{\omega}, \qquad (1)$$

where $\nu$ is the kinematic viscosity. If we assume that the time dependence for velocity is harmonic $\vec{u} \sim e^{i\Omega t}$, where $\Omega$ is the frequency, then the equation for vorticity can be rewritten as:

$$\vec{\omega} = \frac{1}{i\Omega}\left[\left(\vec{\omega}\vec{\nabla}\right)\vec{u} - \left(\vec{u}\vec{\nabla}\right)\vec{\omega} + \nu\vec{\nabla}^2\vec{\omega}\right], \qquad (2)$$

or if the kinematic viscosity is $\nu = 0$ and taking into account the formula for determining vorticity, we obtain:

$$\vec{\omega} = \frac{1}{i\Omega}\left[\left(\left(\vec{\nabla} \times \vec{u}\right)\vec{\nabla}\right)\vec{u} - \left(\vec{u}\vec{\nabla}\right)\left(\vec{\nabla} \times \vec{u}\right)\right]. \qquad (3)$$

Since in this paper we study the vortex behavior of the Poynting vector for electromagnetic plane waves (also with the harmonic time dependence for wave field components $\vec{E}, \vec{H} \sim e^{i\Omega t}$, where $\vec{E}$ and $\vec{H}$ are the electric and magnetic fields of the wave) interacting with 2D photonic crystals, it is logical to introduce vorticity specifically for the Poynting vector by analogy with the case of fluid flow, namely, $\vec{\omega}_P = \vec{\nabla} \times \vec{P} = \frac{1}{2}\vec{\nabla} \times \left(\vec{E} \times \vec{H}^*\right)$. Applying further the well-known vector identity and taking into account Maxwell's equations $\vec{\nabla}\vec{H} = 0$ and $\vec{\nabla}\vec{E} = 0$, we obtain the following expression for the vorticity of the Poynting vector:

$$\vec{\omega}_P = \frac{1}{2}\left[\left(\vec{H}^*\vec{\nabla}\right)\vec{E} - \left(\vec{E}\vec{\nabla}\right)\vec{H}^*\right],$$

or, using Maxwell's equations, one can obtain the following expression for the vorticity of the Poynting vector:

$$\vec{\omega}_P = \frac{ic}{2\Omega}\left[\left(\left(\vec{\nabla} \times \vec{E}^*\right)\vec{\nabla}\right)\vec{E} - \left(\vec{E}\vec{\nabla}\right)\left(\vec{\nabla} \times \vec{E}^*\right)\right] \qquad (4)$$

which, with an accuracy of up to a coefficient, is the same form as the equation for the vorticity of a fluid flow (3). Apparently, there is a connection between the phenomena associated with the vorticity of the fluid flow (3) and the vorticity of the Poynting vector of electromagnetic waves (4) that manifests itself during the interaction of electromagnetic waves with systems that have a complex structure of distribution of dielectric and magnetic permeabilities.

## 3 Reverse energy flows in 2D photonic crystal with dielectric rods

As a second example of reverse energy flows occurring in 2D photonic crystals, let us consider the photonic crystal from the work [21] where dielectric rods with a high refractive index surrounded by air are located at the nodes of a square lattice. The parameters of the photonic crystal are $r = 0.18a$, where $a$ is the distance between two neighboring rods and $r$ is the radius of the rod.

The refractive index of the rods is 3.4. For this configuration of the photonic crystal, a complete photonic band gap can only form for a TE polarized incident plane wave. We will demonstrate this further by calculating the distribution of the Poynting vector $P_x$ component. Just as in the previous case, we will consider a photonic crystal slab consisting of ten layers of dielectric rods.

Based on the results of the work [20], a photonic crystal with a rod radius $r$ = 0.315 µm and a distance between the rods $a$ = 1.75 µm was chosen. These parameters of a photonic crystal allow one to obtain a complete photonic band gap for TE polarized incident plane wave at a wavelength of 5 µm, with a wavelength of 2.78 µm and 3.5 µm both outside the photonic band gap. We plotted the distribution of the Poynting vector $P_x$ component on the right-hand side of the computational domain (Fig. 2) for these three cases (Fig. 8a). From Fig. 8a it is evident that for a TE polarized incident plane wave there is a photonic band gap at a wavelength of 5 µm, but not at wavelengths of 2.78 µm and 3.5 µm. The complete photonic band gap does not exist for TM polarized incident plane wave for all the wavelengths (Fig. 8b). However, the transmission regimes at wavelengths of 5 µm, 3.5 µm and 2.78 µm differ since in the latter case there is a more effective radiation reflection from the photonic crystal slab. The reason for this will be explained below.

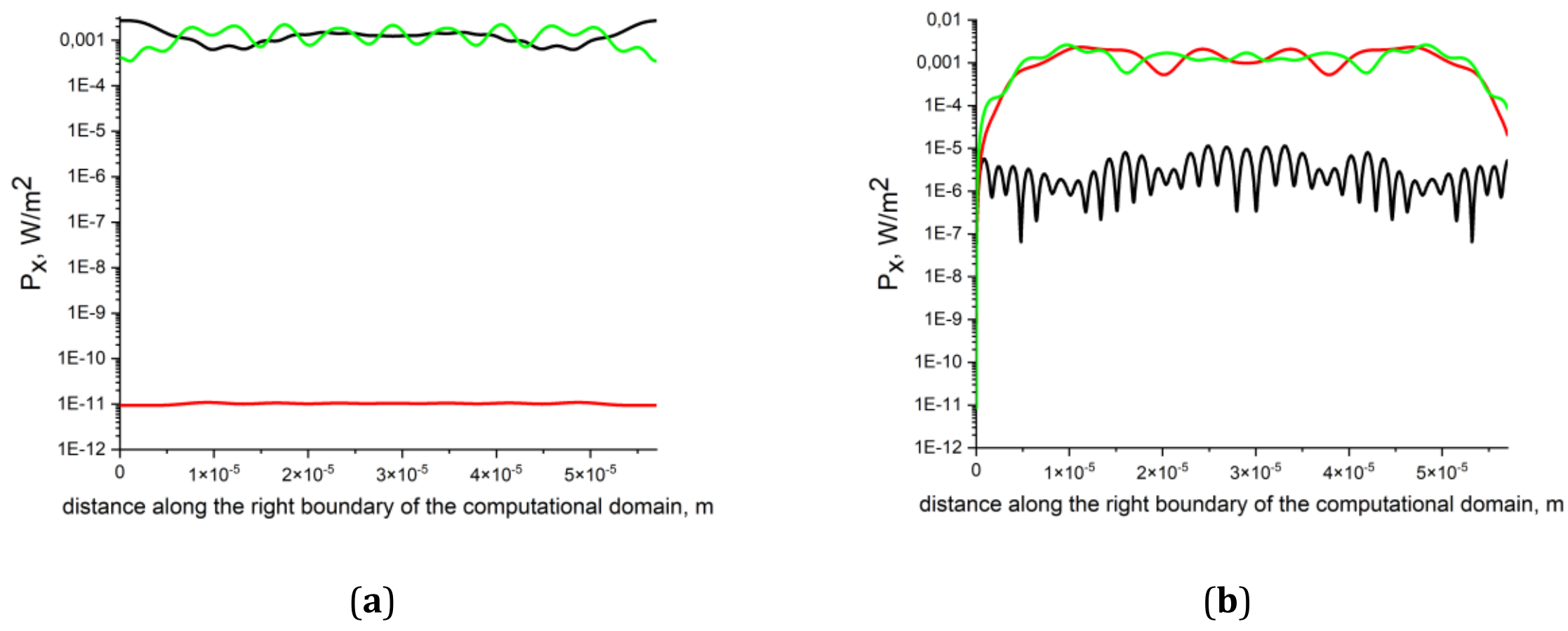

**Figure 8:** (**a**) Distribution of the $P_x$ component of the Poynting vector along the right boundary of the computational domain (Fig. 2) for TE incident polarized plane wave at wavelengths of 5 µm (red), 2.78 µm (black) and 3.5 µm (green); (**b**) the same distributions for TM polarized incident plane wave at wavelengths of 5 µm (red), 2.78 µm (black) and 3.5 µm (green).

The distribution of the Poynting vector $P_x$ component and the Poynting vector streamlines for TE polarized incident plane wave for wavelengths 3.5 µm and 5 µm are shown in Fig. 9.

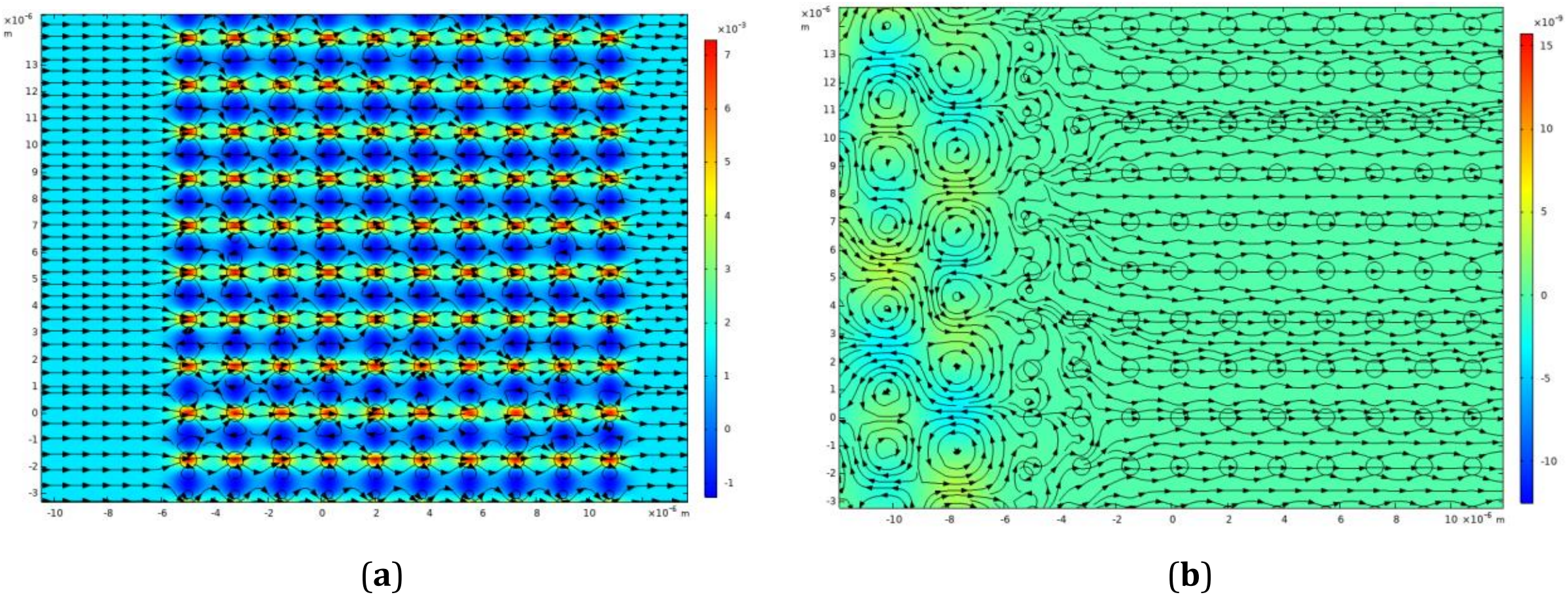

**Figure 9:** (**a**) Streamlines of the Poynting vector $\vec{P}(x, y)$ for TE polarized incident plane wave at wavelength of 2.78 μm on the background of the $P_x$ component distribution; (**b**) streamlines of the Poynting vector $\vec{P}(x, y)$ for the same incident plane wave at wavelength of 5 μm on the background of the $P_x$ component distribution. A plane wave is incident on a photonic crystal slab from the left (Fig. 2).

As it can be seen from Fig. 9a, the reverse energy flows at a wavelength of 2.78 μm are observed only in the areas between the layers of rods of the photonic crystal slab. The distribution of the Poynting vector streamlines shows that the main part of the radiation passes through the rods and there are no vortices outside the photonic crystal slab. For a photonic band gap at a wavelength of 5 μm (Fig.9b), the Poynting vector streamlines form a vortex structure in the region in front of the photonic crystal slab. This leads to a sharp drop in the radiation transmission through the photonic crystal slab (Fig. 8a). For a TM polarized incident plane wave, the transmission of radiation through a photonic crystal slab has large values at wavelengths of 5 μm and 3.5 μm (Fig. 8b).

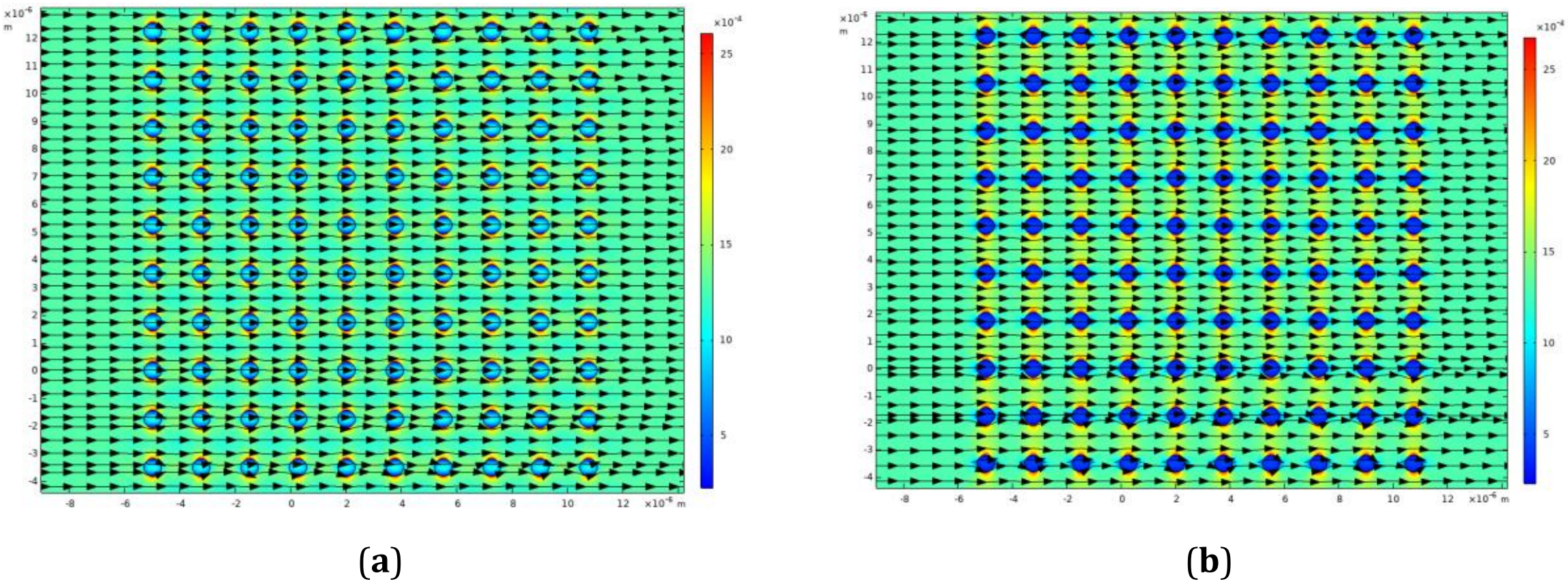

(**a**) (**b**)

**Figure 10:** (**a**) Streamlines of the Poynting vector $\vec{P}(x, y)$ for a TM polarized incident plane wave at wavelength of 3.5 μm on the background of the $P_x$ component distribution; (**b**) streamlines of the Poynting vector $\vec{P}(x, y)$ for the same incident plane wave at a wavelength of 5 μm on the background of the $P_x$ component distribution. A plane wave is incident on a photonic crystal slab from the left (Fig. 2).

This is confirmed by the behavior of the Poynting vector streamlines shown in Fig. 10. From Fig. 10 it is evident that in this case, with strong radiation transmission through the photonic crystal slab, there are no reverse flows or vortex movements of energy.

The situation with radiation transmission changes qualitatively if we consider the incidence of TM polarized plane wave at a wavelength of 2.78 μm (Fig. 8b).

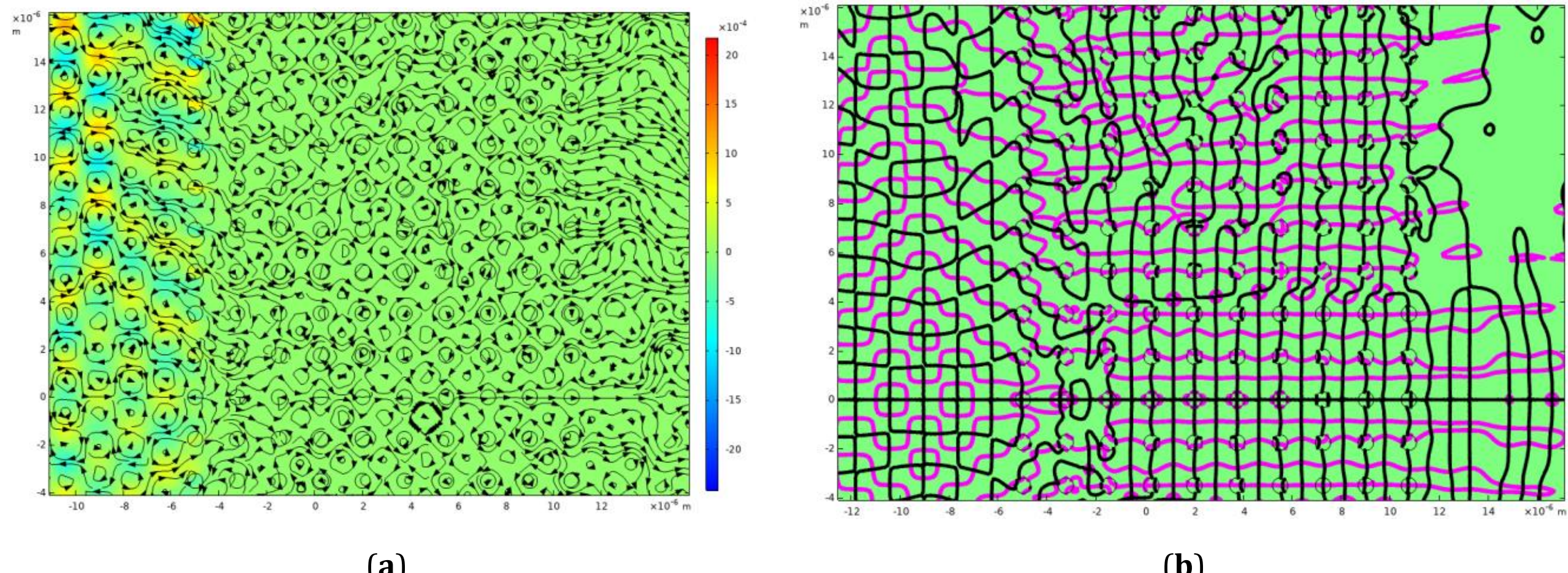

(**a**) (**b**)

**Figure 11:** (**a**) Streamlines of the Poynting vector $\vec{P}(x,y)$ for TM polarized incident plane wave at wavelength of 2.78 μm on the background of the $P_x$ component distribution; (**b**) Curves $P_x(x,y)=0$ (magenta) and $P_y(x,y)=0$ (black) for a TM polarized incident plane wave at wavelength of 2.78 μm (Fig. 8b). The intersection points of these curves are the centers of the Poynting vector vortices.

In this case, the Poynting vector streamlines begin to twist (Fig. 11a) forming reverse energy flows as is in the case of a complete photonic band gap for a TE polarized incident plane wave (Fig. 8a). This behavior of the energy flow is caused by Poynting vector vortices, whose centers lie at the intersection of curves $P_x(x,y)=P_y(x,y)=0$ (Fig.11b). This, in turn, explains the drop in radiation transmission at a wavelength of 2.78 μm (Fig. 8b).

Thus, it can be argued that the formation of photonic band gaps in 2D photonic crystals is always accompanied by the emergence of vortex motions of the Poynting vector for reflected and transmitted radiation. This in turn leads to the formation of reverse flows of electromagnetic field energy in certain areas of the photonic crystal cross-section.

## 4 Conclusion

In this paper, we considered the transmission of TM and TE polarized plane waves through photonic crystal slabs whose optical properties were obtained in previous studies. Calculations confirmed the presence of photonic band gaps in the required spectral regions. By analogy with hydrodynamics, in which the flow of fluids is characterized by velocity vectors, we examined the behavior of the Poynting vector for propagating electromagnetic radiation in photonic crystal slabs. It turned out that, as is in the case of a fluid encountering obstacles, vortices—regions of swirling flow—can arise in the distribution of the Poynting vector, which in turn generate reverse flows of electromagnetic radiation energy. The decrease in the transmission of electromagnetic radiation through photonic crystal slabs (photonic band gaps) is characterized by the formation of reverse energy flows and Poynting vector vortices. Moreover, this is true for different types of two-dimensional photonic crystals. Therefore, while the physics and governing equations differ (Maxwell's equations for electromagnetic waves vs. Navier-Stokes equations for fluid flow), the conceptual framework of vector fields, vortex formation, and reverse flow provides a meaningful analogy. This analogy can inspire new ways to understand, visualize, and model complex electromagnetic phenomena by borrowing insights and mathematical tools from classical fluid dynamics.

**Acknowledgement:** The author would like to thank Dr. Alexey Kosolapov and Dmitry Komissarov for their help in designing the drawings for this article and useful discussions.

**Funding Statement:** The author received no specific funding for this study.

**Availability of Data and Materials:** The data that support the findings of this study are available from the Corresponding Author, A. P., upon reasonable request.

**Ethics Approval:** Not applicable.

**Conflicts of Interest:** The author declares no conflicts of interest to report regarding the present study.

## References

.